\documentclass{article}

    \PassOptionsToPackage{numbers, compress}{natbib}

    \usepackage[preprint]{neurips_2021}

\usepackage[utf8]{inputenc} 
\usepackage[T1]{fontenc}    
\usepackage{hyperref}       
\usepackage{url}            
\usepackage{booktabs}       
\usepackage{amsfonts}       
\usepackage{nicefrac}       
\usepackage{microtype}      
\usepackage{graphicx}
\usepackage{xcolor}
\usepackage{float}
\usepackage[export]{adjustbox}

\usepackage[english]{babel}
\usepackage{blindtext}

\title{Using Deep Learning Sequence Models to Identify SARS-CoV-2 Divergence}

\author{

    Yanyi Ding \\
    Healthcare Analytics \& IT \\
    Carnegie Mellon University \\
    Pittsburgh, PA 15213 \\
    \texttt{yanyid@andrew.cmu.edu} \\
    
    \And
    
    Zhiyi Kuang \\
    Computational Finance \\
    Carnegie Mellon University \\
    Pittsburgh, PA 15213 \\
    \texttt{zkuang@andrew.cmu.edu} \\
    
    \And
    
    Yuxin Pei \\
    Computer Science \\
    Carnegie Mellon University \\
    Pittsburgh, PA 15213 \\
    \texttt{yuxinp@andrew.cmu.edu} \\
    
    \And
    
    Jeff Tan \\
    Computer Science \\
    Carnegie Mellon University \\
    Pittsburgh, PA 15213 \\
    \texttt{jefftan@andrew.cmu.edu} \\
    
    \And
    
    Ziyu Zhang \\
    Computer Science \\
    Carnegie Mellon University \\
    Pittsburgh, PA 15213 \\
    \texttt{ziyuzhan@andrew.cmu.edu} \\
    
    \And
    
    Joseph Konan (TA) \\
    Electrical \& Computer Eng. \\
    Carnegie Mellon University \\
    Pittsburgh, PA 15213 \\
    \texttt{jkonan@andrew.cmu.edu} \\

}
\begin{document}

\maketitle

\begin{abstract}
SARS-CoV-2 is an upper respiratory system RNA virus that has caused over 3 million deaths and infecting over 150 million worldwide as of May 2021. With thousands of strains sequenced to date, SARS-CoV-2 mutations pose significant challenges to scientists on keeping pace with vaccine development and public health measures. Therefore, an efficient method of identifying the divergence of lab samples from patients would greatly aid the documentation of SARS-CoV-2 genomics. In this study, we propose a neural network model that leverages recurrent and convolutional units to directly take in amino acid sequences of spike proteins and classify corresponding clades. We also compared our model's performance with Bidirectional Encoder Representations from Transformers (BERT) pre-trained on protein database. Our approach has the potential of providing a more computationally efficient alternative to current homology based intra-species differentiation.

\end{abstract}

\section{Introduction}
\label{sec:intro}
SARS-CoV-2 is an RNA virus that has caused over 150 million infections and 3 million deaths worldwide as part of the COVID-19 pandemic. SARS-CoV-2 spreads mainly through the air when people breathe, cough, sneeze, or speak in close proximity, and has caused widespread social and economic disruption. As SARS-CoV-2 constantly mutates during replication of genomic information \cite{Pachetti}, the mutation rate is one of the most crucial parameters to assess the risk of this infectious disease~\cite{Pathan}. For example, Lineage B.1.617 has participated in an upsurge of COVID-19 in India throughout early 2021, with multiple mutations that may result in better transmissibility \cite{Cherian}.

Analyzing genomic divergence, mutations, and clade across viral protein sequences allows scientists to understand where specific strains are concentrated, the virulence of different clades, and their resistance to vaccines and treatments. In particular, the SARS-CoV-2 spike protein plays a key role in cell membrane fusion, and is commonly used as a target for vaccine and therapeutic development \cite{Du}. However, there is a vast landscape of strains and mutations for SARS-CoV-2 \cite{Lauring}, with the constant birth and death of viral lineages and the movement of existing lineages between populations. This genetic diversity can make it difficult to maintain a robust classification system for SARS-CoV-2 that is flexible enough to accommodate new viral diversity, while ensuring that existing genomes are classified into descriptive categories \cite{Rambaut}.

We aim to build and analyze a sequence-to-one recurrent classification model that can be trained to classify viral protein sequences by their divergence, mutations, clade, and potentially any other classification system developed in the future. Recurrent neural architectures such as long short-term memory (LSTM) \cite{Hochreiter} and gated recurrent units (GRU) \cite{Cho} are able to model long-term dependencies within the input sequence, and are therefore well suited for modelling the functions and characteristics of protein sequences \cite{Liu} such as the SARS-CoV-2 spike protein. As deep neural networks can be trained to output arbitrary classifications based on how the training data is labelled, a well-trained deep learning model could adapt to new classification systems through the use of transfer learning methods. In addition, deep neural networks are also a scalable solution to the problem of sequence classification. Once the model is trained, inference can occur without human intervention on large datasets of protein sequences, without the need to search through large databases of protein sequences.
\section{Related Work}
\label{sec:related_work}
Since the outbreak of the COVID-19 pandemic, an extensive amount of research has been done to understand the evolution of SARS-CoV-2. \cite{Domenico} provide a phylodynamic and phylogeographic analysis and discovered evidence supporting the origin of SARS-CoV-2. With the genome sequences of SARS-CoV-2 and the Bat SARS-like Coronavirus from GeneBank, they employed a flexible uncorrelated relaxed molecular clock model in the Bayesian phylogenetic analysis. A Maximum Clade Credibility tree was then built from the tree posterior distribution. They found sufficient phylogenetic signals identifying the most probable origin of SARS-CoV-2. Our goal is to reproduce similar results with a deep learning approach.

Many researchers have also tackled the SARS-CoV-2 challenge using machine learning. CRISPR-based viral detection systems have been designed with high sensitivity and speed \cite{Wang}. There are also neural networks capable of learning motif embeddings and discriminating viral genomes without requiring alignment. Tools such as DeepMicrobes can classify metagenomic samples into bacterial genuses and estimate abundance using a convolutional neural network model combined with bidirectional LSTM layers \cite{Liang}. Similarly, for viral classification, DeepVirFinder and Viral Miner were able to learn k-mer patterns using the advantage of the convolutional neural network's location invariance, and classify contigs of 300 bp with AUC above 93\%. \cite{Ren, Tampuu}. Neural network models provide an improvement to traditional homology based alignment such as BLAST because it does not require the time and CPU needed to search through a vast database and the potential for previously unidentified sequences.

While there are many neural networks that can differentiate microbes or viruses among each other, not much research has been done for intra-species divergence differentiation. Nonetheless, we believe that existing research has shed light on how we could approach building deep learning architectures for measuring divergence. We were able to find similar attempts on mutation rate prediction \cite{Pathan} and genetic sequence generation \cite{Lim}, which inform our baseline specification.

\cite{Lim} developed a sequence-to-sequence bidirectional LSTM model EvoLSTM that takes in an ancestral genomic sequence, calculates a probability distribution over possible descendant characters at each position, and outputs a descendant genomic sequence. EvoLSTM consists of connected LSTM-based encoder and decoder networks, following the approach in \cite{Sutskever}. The encoder network consists of two LSTMs, which look at the same input sequence but in opposite directions. The decoder network consists of an LSTM followed by a fully connected network, which makes use of a multi-layer perceptron and a softmax activation to produce a probability distribution over possible descendant nucleotides for each position. According to \cite{Lim}, the LSTM provides information to each position about the evolutionary event that took place at the previous position, which provides more context in its predictions. EvoLSTM is trained on entire whole-genome primate alignments and demonstrates its ability to capture long distance mutational context dependencies. 

\cite{Pathan} created an RNN-based LSTM model to predict the mutation rate of the virus in patients affected by COVID-19. Specifically, they ``calculated the base substitution mutation rates from analyzing the genomic sequence of 3408 patients from different countries, focusing mainly on mutations that have developed freely on different dates.'' \cite{Pathan}'s baseline LSTM model consists of one LSTM layer, following by two dense layers and ReLU activation, interleaved with 0.25 dropout layers, and a final dense output layer with the Adam optimizer. It gives Root Mean Square Error (RMSE) of 0.06 in testing and 0.04 in training. The trained model can be used to predict future patients' mutation rate. 

After extensive literature search, we are not aware of a deep-learning based model that targets the same objective (predicting the divergence) using protein sequences. All previous work that studies divergence are based on biological methods (e.g. posterior probability). Therefore, we are currently only comparing our result with ground truth data.
\section{Dataset}

The dataset of interest for this study has only become available in recent years. For this reason, there are multiple challenges and uncertainties when working with this data. We discuss the data source, dataset class, and data loader below. The following figure summarizes our data processing methodology:


\begin{figure}[!ht]
    \centering
    \includegraphics[width=0.8\textwidth]{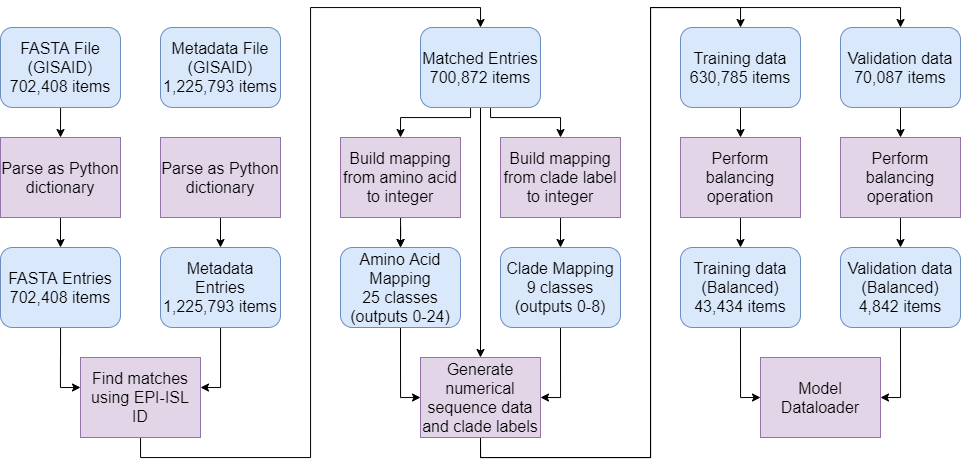}
    \caption{Data processing method}
    \label{fig:DataPreprocess}
\end{figure}

\subsection{Source}

The dataset for this problem is protein sequences that encode the spike protein of SARS-CoV-2, and with metadata describing the mutations, clade, and Pango lineage of each sequence, both obtained from GISAID \cite{Elbe, Shu}. Using protein sequences provides the advantage of dismissing the insignificant effect of silent mutations at the DNA level, which our model would otherwise have to learn if we used raw genomic data. The encoded protein sequences are in FASTA format, and each entry has a title line with an EPI-ISL accession identifier. The encoded metadata is in TSV format, and each entry also has an EPI-ISL accession identifier used to match metadata entries with FASTA entries. Most protein sequences have about 1274 amino acids, with some minor differences in length due to insertion or deletion mutations. More details on this dataset, including example entries and the amino, clade, and Pango lineage mappings, are included in Appendix I.

Input format: FASTA (see appendix for more detailed example)
\begin{verbatim}
    >title line
    MXWXXX......BCG* 
    (About 1274 amino acid sequence from spike protein,
    asterisk indicates ending of a sequence, X indicates unknown amino acid.)
\end{verbatim}

\subsection{Dataset Class}

We begin by reading the FASTA protein sequences and TSV metadata,  parsing them into Python dictionaries indexed by EPI-ISL ID. Then, we iterate through the protein sequence and metadata dictionaries, building a list of EPI-ISL IDs that match between both files. Once the correspondence is made, we create a mapping from amino acids and clade labels to integers. We also use these mappings to convert each of our matched data points to a numerical representation. Finally, we set aside every 10th entry as validation data and keep the remaining entries as training data. Like clades, we perform a similar process for Pango lineage, whose label has 1271 classes compared to the clade's 9 classes. The result is a set of training and validation instances whose pairs consist of tokenized-numeric versions of the input protein sequences that map to clade class labels.

\subsection{Data Loading}

One of the issues with our dataset above is that certain output classes dominate the dataset. Shown below is a histogram of clade and Pango lineage label frequencies. In both the training and validation data, clades 1, 2, 3, 4, and 0 appeared much more frequently than clades 5, 6, 7, and 8, with clade 1 appearing over 30x more often than clade 5 in both datasets. In addition, out of the 1271 Pango lineages in the dataset, the five most common lineages account for over 45\% of the entire dataset, and many of the least common lineages occur fewer than ten times total across both datasets. For this reason, we use clade labels instead of Pango lineage labels as our classification targets for this task.

\begin{figure}[H]
    \hrule\vspace{0.1in}\centering
    \includegraphics[scale=0.45]{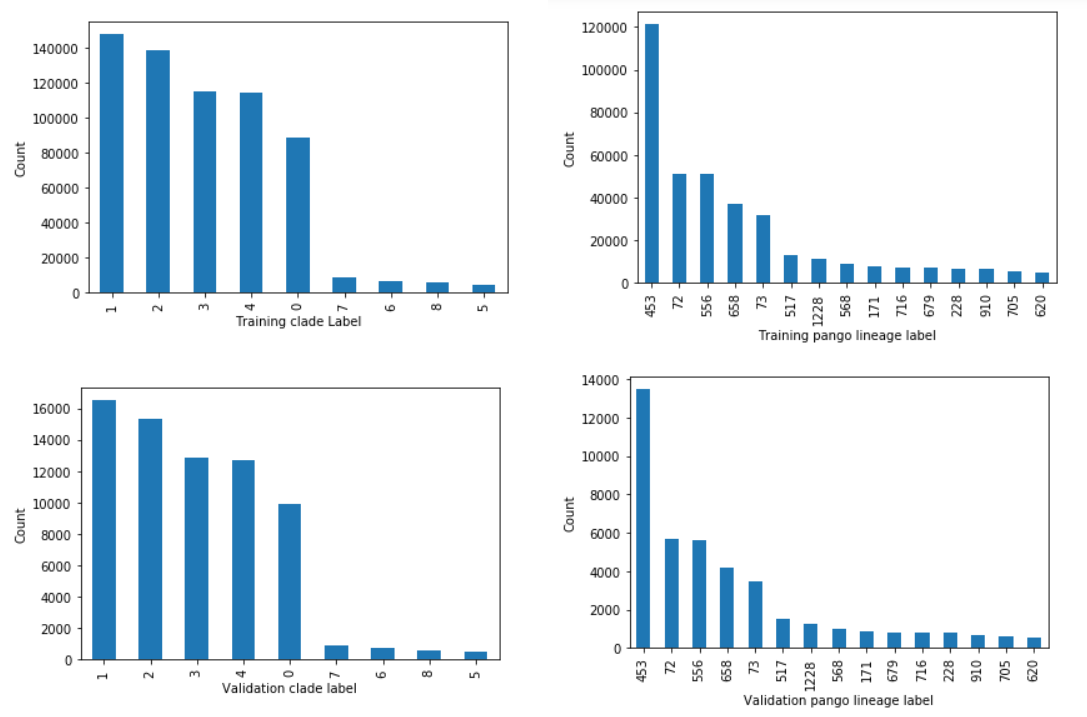}
    \vspace{0.1in}\hrule
    \caption{Label frequency for clade and Pango lineage in full training and evaluation datasets}
    \label{fig:LabelFrequency}
\end{figure}

To further address this issue, we balanced the data by downsampling to ensure similar item counts for each output class. Specifically, we use RandomUnderSampler from the imbalanced-learn Python package\footnote{\url{imbalanced-learn.org/stable/references/generated/imblearn.under_sampling.RandomUnderSampler.html}}. First, we determine how many times the least common output class occurred, then we select a subset of the data for all other output classes so that each class appears equally often.

This balancing step helped resolve two issues. First, it improved our training time by removing redundant data entries. Our original dataset was very large, taking over 2 hours to train per epoch on hardware provided by Google Colab, but training on the reduced dataset was over 14x faster. In addition, as some output classes dominated others in the original dataset, our model was just learning to output the most common classes, instead of learning a meaningful representation of the input data. Balancing the dataset prevents the model from achieving good performance simply by memorizing and outputting the most common class.

\section{Model Description and Results}
In this section, we present the various model architectures that we tried. Although none of our models produced particularly good results, we aim to analyze why we think our models are performing poorly, and what changes would be needed to improve their results.

We formulate the problem as a classification problem. The input to our model at each timestep is 
$$x_t = s_t$$
where $s_t$ is the standard code for the $t^{th}$ peptide on the spike protein sequence of the sample. The output of our model is the predicted clade of the given sample, represented as a softmax probability vector where the neuron with maximal probability indicates the most probable clade of the given sample. The number of output classes is 9. Our loss function is cross entropy loss, and the model is being evaluated against a pre-sampled, ground-truth validation dataset. The evaluation metrics and the workflow are as follows:


\begin{figure}[H]
    \hrule\vspace{0.1in}\centering
    \includegraphics[scale=0.45]{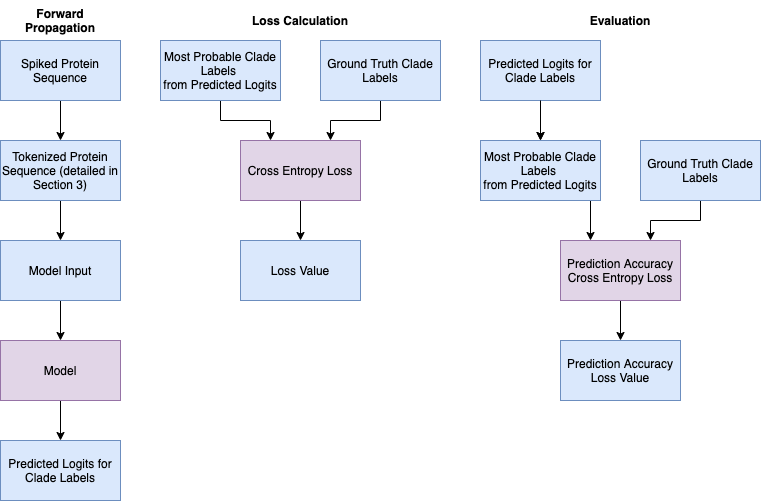}
    \vspace{0.1in}
    \hrule
    \caption{Workflow}
    \label{fig:Flow}
\end{figure}

The following is the pseudocode for training. 
\begin{verbatim}
    Match between sequence and metadata files using EPI-ISL ID.
    Sample 1/10 of all sequences as the verification data.
    Transform to numerical data type.
    For every epoch:
      For every batch:
        CNN.forward()
        LSTM.forward()
        Linear.forward()
        CNN.backward()
        LSTM.backward()
        Linear.backward()
    Save model
    Output training accuracy
    Run current model on verification data
    Output verification accuracy
\end{verbatim}
Our desired end result is an end-to-end system, therefore, we'll add necessary data preprocessing in the inference code, per advice from collaborators in the medical field.

\subsection{Classification Objective}
Given a set of inputs $\{\mathbf{x}_i\}_{i=1}^N, \mathbf{x}_i\in\mathbb{R}^M$ and a set of corresponding labels $\{y_i\}_{i=1}^N$, where $M$ is the maximum length of a spike protein sequence and $N$ is the size of the dataset, the softmax layer outputs the predicted probability for the $j$-th class for the sample vector $\mathbf{x}_i$ and weight vector $\mathbf{w}$ as
\[\mathbb{P}(y_i=j\mid\mathbf{x}_i)=\frac{e^{\mathbf{x}_i^T\mathbf{w}_j}}{\sum_{k=1}^Ne^{\mathbf{x}_i^T\mathbf{w}_k}}\]
We output the class with the highest predicted probability. Denote $\hat{\mathbf{y}}$ as the output of the network, then the cross-entropy loss is
\[\ell(\mathbf{w})=-\sum_{i=1}^N\sum_{k=1}^9y_k^{(i)}\log\left(\hat{y}_k^{(i)}\right)\]
We evaluate model performance using cross-entropy loss and prediction accuracy. Given a set of predicted labels $\hat{\textbf{y}}=\{\hat{y_i}\}_{i=1}^N$ and ground truth labels $\textbf{y}=\{y_i\}_{i=1}^N$, the prediction accuracy is $$Acc(\hat{\textbf{y}},\textbf{y}) = \frac{\sum_{i=1}^N\textbf{1}_{\{y_i = \hat{y}_i\}}}{N}$$

All of the runs described below were implemented using PyTorch. We used the Adam optimizer with learning rate 0.002 and weight decay 0.000005, and the ReduceLROnPlateau learning rate scheduler with factor 0.5 and patience 2. Our batch size was set to be as large as possible without encountering out-of-memory errors, typically we used batch size 128. 

\subsection{Model Design and Motivation}

This project is particularly challenging because there are no reference models to date in the literature that specifically model SARS-CoV-2 protein sequences. We postulate that the complexity of these models are no more complex than language modeling, whose hw3p2 performance motivates Specification A. There is a trade-off between having complex models that are difficult to iterate on versus smaller models that iterate quickly. In the coming sections, each of these specifications is expounded and their results are explained.

Inspired by the findings in \cite{Pathan}, we use an LSTM-based architecture \cite{Hochreiter} as our baseline model for this problem, with convolutional layers as a feature extractor for sequential data. Convolutional layers are able to extract features by applying a non-linearity on an affine combination of inputs from each time window for each feature \cite{Keren}. Furthermore, LSTMs are well-suited for processing sequential data, as LSTM cells are able to remember values across arbitrary intervals of the input sequence. In theory, this makes them well-suited for classifying protein sequences, as the presence of different combinations of mutations may have complex effects on how sequences get classified. For example, Lineage B.1.671 is identified by a combination of at least 6 signature mutations, 3 of which occur commonly in other globally circulating lineages \cite{Cherian}. A successful classifier would need to determine whether this combination of mutations is present to correctly classify these sequences.

\subsection{Baseline Model: Specification A}

Our baseline model consists of one convolutional layer with 128 output channels and a kernel size of 3, with BatchNorm and ReLU applied. Next, we have three bi-LSTM layers with 256 hidden states in each direction, and two fully connected layers using ReLU activation with 512 and 256 neurons respectively, as well as BatchNorm and Dropout=0.2. We use a log softmax layer as our final output. We selected this baseline based on findings in \cite{Pathan}, as well as our own experiences in hw3p2.

\begin{figure}[H]
    \hrule\vspace{0.1in}\centering
    \includegraphics[scale=0.45]{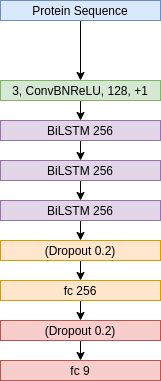}
    \vspace{0.1in}
    \hrule
    \caption{Model Architecture for Specification A}
    \label{fig:ArchitectureBaseline}
\end{figure}

\begin{verbatim}
==========================================================================
                      Kernel Shape       Output Shape     Params Mult-Adds
Layer                                                                   
0_conv.Conv1d_0        [1, 128, 3]  [ 32,  128, 1274]      384.0  489.216k
1_conv.BatchNorm1d_1         [128]  [ 32,  128, 1274]      256.0     128.0
2_conv.ReLU_2                    -  [ 32,  128, 1274]          -         -
3_lstm                           -  [1274,  32,  512]  3.944448M  3.93216M
4_linear.Dropout_0               -  [1274,  32,  512]          -         -
5_linear.Linear_1       [512, 256]  [1274,  32,  256]   131.328k  131.072k
6_linear.Dropout_2               -  [1274,  32,  256]          -         -
7_linear.Linear_3         [256, 9]  [1274,  32,    9]     2.313k    2.304k
8_linear.LogSoftmax_4            -  [1274,  32,    9]          -         -
--------------------------------------------------------------------------
Total params          4.078729M            Non-trainable params        0.0
Trainable params      4.078729M            Mult-Adds              4.55488M
==========================================================================
\end{verbatim}
\centerline{Table 1: Model Metadata for Specification A}


\subsubsection{Baseline Results on Unbalanced Data}

We initially used our baseline model on the unbalanced dataset, and our results are as follows:
\begin{figure}[H]
    \hrule\vspace{0.1in}\centering
    \includegraphics[scale=0.45]{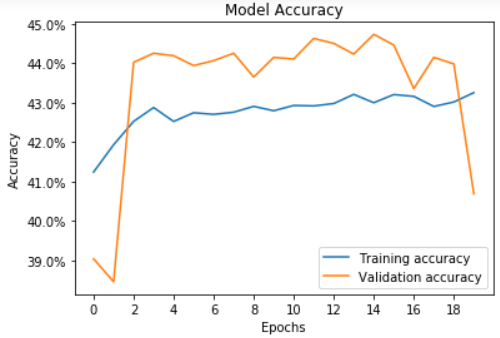}
    \includegraphics[scale=0.45]{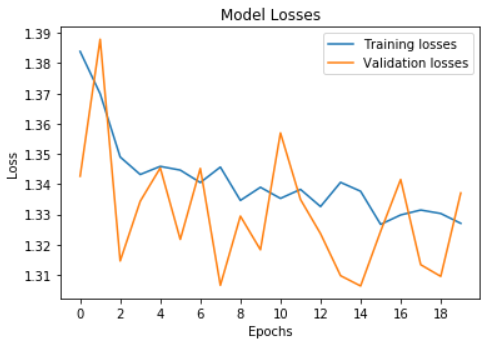}
    \vspace{0.1in}\hrule
    \caption{Accuracy and loss for baseline model}
    \label{fig:AccuracyLossBaseline}
\end{figure}
We also plotted a histogram of model output frequencies vs. label frequencies, as well as a confusion matrix describing the classification error of the model:
\begin{center}
    \hrule\vspace{0.1in}
    \includegraphics[scale=0.45]{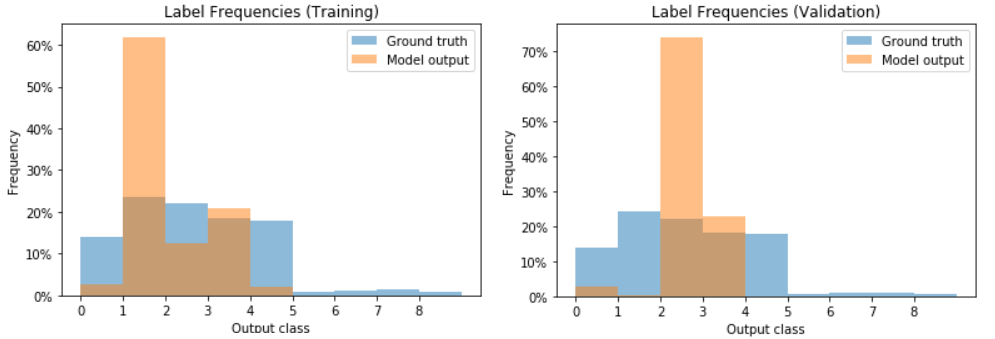}
\end{center}
\begin{figure}[H]
    \centering
    \includegraphics[scale=0.45]{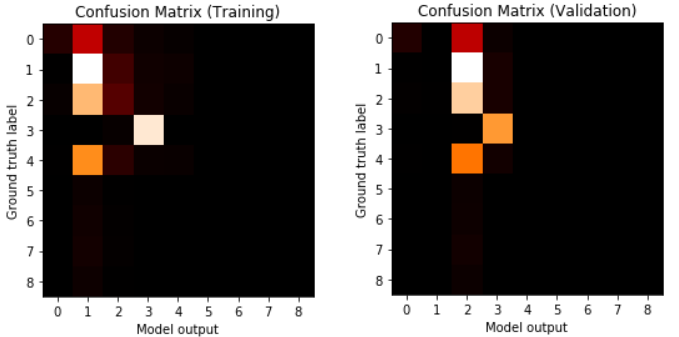}
    \vspace{0.1in}\hrule
    \caption{Output label frequencies and confusion matrix for baseline model}
    \label{fig:FrequencyConfusionBaseline}
\end{figure}

Although the model seemed to successfully identify clade 3, it was unable to identify most of the remaining clades, and seemed to pick an arbitrary output class for those inputs (clade 1 in training, clade 2 in evaluation). The model also did not output any predictions for clades 5-8. We hypothesize that because clades 1 and 2 are the most common output classes, outputting clade 1 or clade 2 at all times maximizes the model's success probability when it is not able to distinguish the clade of the input sequence. In an attempt to resolve this issue, we introduced a data balancing step.

\subsubsection{Baseline Results on Balanced Data}

After implementing the dataset balancing step, our results on the same model as above are as follows:
\begin{figure}[H]
    \hrule\vspace{0.1in}\centering
    \includegraphics[scale=0.45]{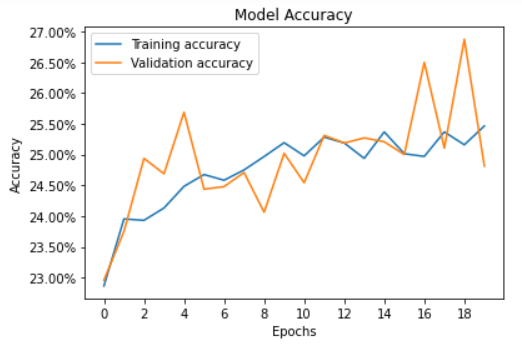}
    \includegraphics[scale=0.45]{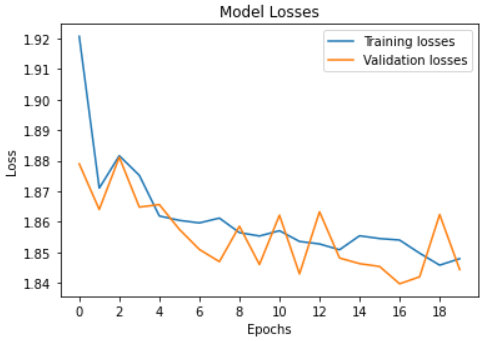}
    \vspace{0.1in}\hrule
    \caption{Accuracy and loss for baseline model with balanced dataset}
    \label{fig:AccuracyLossBaselineBalanced}
\end{figure}

We also plotted a histogram of model output frequencies vs. label frequencies, as well as a confusion matrix describing the classification error of the model:
\begin{center}
    \hrule\vspace{0.1in}
    \includegraphics[scale=0.45]{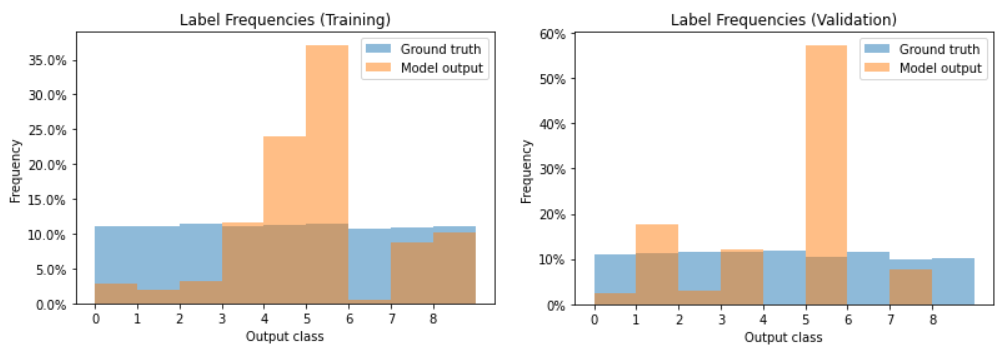}
\end{center}
\begin{figure}[H]
    \centering
    \includegraphics[scale=0.45]{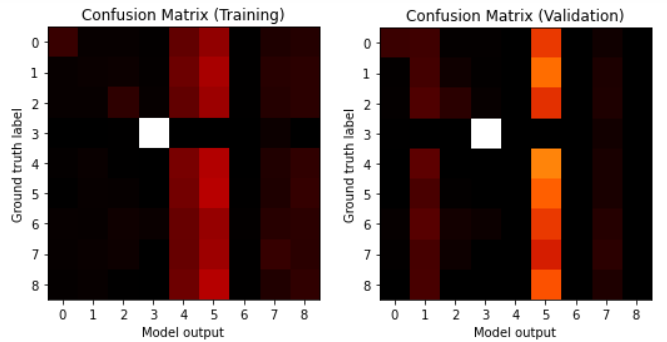}
    \vspace{0.1in}\hrule
    \caption{Output label frequencies and confusion matrix for baseline model with balanced dataset}
    \label{fig:FrequencyConfusionBaselineBalanced}
\end{figure}

Even though our performance was higher on the unbalanced dataset, this is because there is no longer as much advantage in outputting the most common output class at all times. Our model still has similar issues as before, as it is able to reliably identify clade 3, but seems unable to identify any of the other clades, and defaults to outputting a single label in the majority of cases. Therefore, we believe that our model's internal representation of the input sequence is not expressive enough to distinguish between the different clade labels.

\subsubsection{Hyperparameter Search for Baseline Model}
\begin{center}
 \begin{tabular}{||c c c c c c||} 
 \hline
 Batch Size & Train Set Size & Data Augmentation & Label & Max Validation Acc & AUC \\ [0.5ex] 
 \hline\hline
 32  & 2400 & None & mutation & 44.4\% & 0.55 \\ 
 \hline
 1 & 2400 & random\_replace(40) & mutation & 42.2\% & 0.49\\
 \hline
 1 & 2400 & random\_replace(40) & clade & 28.5\% & 0.18\\
 \hline
 128 & 40k+ & None & pango lineage & 26.3\% & 0.16\\
 \hline
 128 & 40k+ & None & clade & 40\% & 0.13 \\
 \hline
 128 & 40k+ Balanced & random\_replace(40) & clade & 14\% & 0.12 \\ [1ex] 
 \hline
\end{tabular}
\end{center}
Note that AUC is calculated with classes binarizer and scaling each class's weight by the inverse of the proportion in the data.

\clearpage

\subsection{Pyramidal LSTM: Specification B}

Although LSTM architectures avoid the issue of vanishing gradients that affects traditional RNN architectures \cite{Sherstinsky}, our baseline model remains very difficult to train as evidenced by our gradient flow analysis in Section 5.1. Inspired by \cite{Chan}, we propose a pyramidal LSTM architecture that halves the time resolution of the input at each layer, reducing the length of our data sequence and hopefully making it easier to train our network.

In a typical bi-LSTM network, the output at the $i$-th time step from the $j$-th layer is computed as:
$$h_i^j = BLSTM(h_{i-1}^j,h_i^{j-1})$$
In a pyramidal bi-LSTM network, we concatenate the outputs at consecutive steps of each layer, as:
$$h_i^j = pBLSTM(h_{i-1}^j,[h_{2i}^{j-1},h_{2i+1}^{j-1}])$$

In our model, following the approach of \cite{Chan}, we stack three pyramidal bi-LSTM layers on top of a regular bi-LSTM layer at the bottom. Our model description is as follows:

\begin{figure}[H]
    \hrule\vspace{0.1in}\centering
    \includegraphics[scale=0.45]{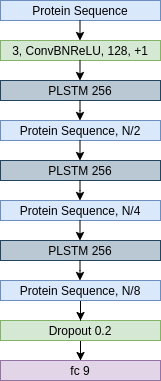}
    \vspace{0.1in}
    \hrule
    \caption{Model Architecture for Specification B}
    \label{fig:ArchitecturePyramidal}
\end{figure}

\begin{verbatim}
=============================================================================
                        Kernel Shape       Output Shape    Params   Mult-Adds
Layer                                                                      
0_embedding                 [32, 25]  [1274,  32,   32]     800.0       800.0
1_conv.Conv1d_0         [32, 128, 3]  [  32, 128, 1274]   12.288k  15.654912M
2_conv.BatchNorm1d_1           [128]  [  32, 128, 1274]     256.0       128.0
3_conv.ReLU_2                      -  [  32, 128, 1274]         -           -
4_lstm                             -  [1274,  32,  256]  264.192k    262.144k
5_pblstm1.LSTM_blstm               -  [ 637,  32,  256]  657.408k     655.36k
6_pblstm2.LSTM_blstm               -  [ 318,  32,  256]  657.408k     655.36k
7_pblstm3.LSTM_blstm               -  [ 159,  32,  256]  657.408k     655.36k
8_linear.Linear_0         [256, 128]  [ 159,  32,  128]   32.896k     32.768k
9_linear.Dropout_1                 -  [ 159,  32,  128]         -           -
10_linear.Linear_2          [128, 9]  [ 159,  32,    9]    1.161k      1.152k
11_linear.LogSoftmax_3             -  [ 159,  32,    9]         -           -
-----------------------------------------------------------------------------
Total params           2.283817M             Non-trainable params         0.0
Trainable params       2.283817M             Mult-Adds             17.917984M
=============================================================================
\end{verbatim}

\centerline{ Table 2: Model Metadata for Specification B}

Our results are as follows:
\begin{figure}[H]
    \hrule\vspace{0.1in}\centering
    \includegraphics[scale=0.45]{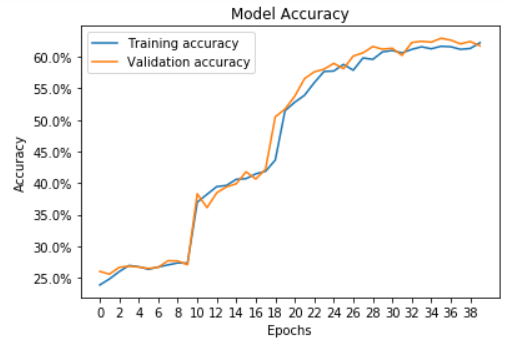}
    \includegraphics[scale=0.45]{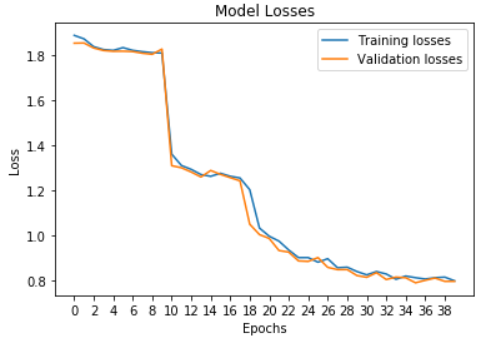}
    \vspace{0.1in}\hrule
    \caption{Accuracy and loss for pyramidal LSTM model}
    \label{fig:AccuracyLossPyramidal}
\end{figure}

We also plotted a histogram of model output frequencies vs. label frequencies, as well as a confusion matrix describing the classification error of the model:
\begin{center}
    \hrule\vspace{0.1in}
    \includegraphics[scale=0.45]{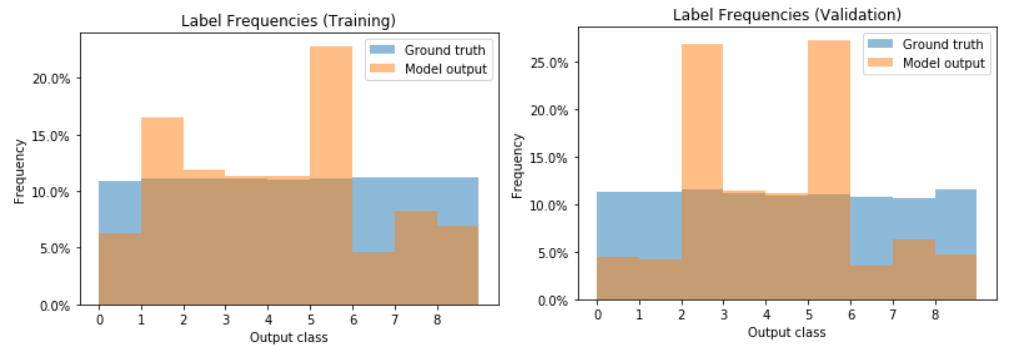}
\end{center}
\begin{figure}[H]
    \centering
    \includegraphics[scale=0.45]{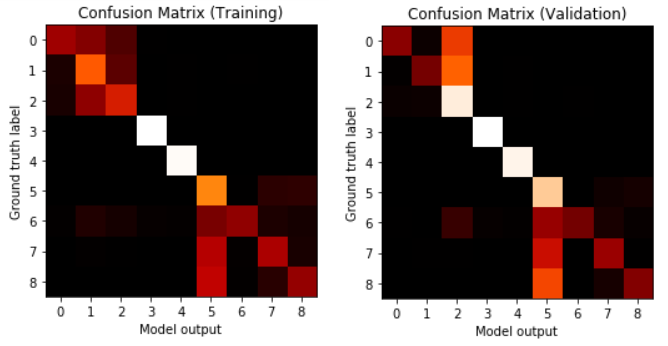}
    \vspace{0.1in}\hrule
    \caption{Output label frequencies and confusion matrix for pyramidal LSTM model}
    \label{fig:FrequencyConfusionPyramidal}
\end{figure}

This model seems to have much better performance, and has learned how to identify clades 2-5 with good accuracy, as well as partly how to identify clades 0-1 and 6-8. The model has also learned to group together clades 0-2 and clades 5-8, which reflects that these clades are relatively similar to each other according to the T-SNE embedding analysis discussed in Section 5.2. When the model is uncertain about how to label a particular input, it seems to default to either clade 2 or clade 5, which is consistent with the behavior of previous models. In total, this model achieves better performance and converges much more quickly than any of the standard LSTM-based models. We hypothesize that this is because it shortens the input sequences to a length that is reasonable for LSTM to handle.

\subsubsection{Hyperparameter Search for Specification B}
\begin{center}
    \begin{tabular}{||c c c c c c||} 
    \hline
    Initial LR & Weight Decay & Encoder & Classifier & Training & Validation \\
    & (Adam) & Dropout & Dropout & Acc & Acc \\ [0.5ex] 
    \hline\hline
    2e-3 & 0 & 0 & 0.2 & 54.5\% & 53.7\% \\ 
    \hline
    2e-3 & 0 & 0 & 0.3 & 55.3\% & 54.6\% \\ 
    \hline
    3e-3 & 5e-5 & 0 & 0.2 & 25.2\% & 24.8\% \\
    \hline
    2e-3 & 1e-6 & 0 & 0.2 & 49.7\% & 49.1\% \\
    \hline
    2e-3 & 1e-7 & 0 & 0.2 & 55.2\% & 54.9\% \\
    \hline
    2e-3 & 1e-8 & 0.15 & 0.2 & 62.3\% & 59.7\% \\
    \hline
    2e-3 & 1e-8 & 0.3 & 0.2 & 47.6\% & 45.9\% \\
    \hline
    2e-3 & 1e-8 & 0.1 & 0.2 & 64.1\% & 63.8\% \\
    \hline
    \end{tabular}
\end{center}
The training uses an ReduceLROnPlateau scheduler with factor 0.5 if the optimizer does not have weight decay, and 0.7 if the optimizer has weight decay. Training was warm restarted with a bigger LR as needed.

\subsection{ Bert Embedding + Hydrophilicity Encoding: Specification C}

Inspired by \cite{Li}, we also experimented on Bert embedding schemes. Instead of simply encoding each amino acid as an index that has no specific biological meaning, we concatenate the Bert embedding pre-trained on protein sequences \cite{AlQuraishi} with the amino acid hydrophilicity encoding. Hydrophilicity refers to how soluble an amino acid is in water, and it is an important property of amino acids with significant effect on protein function \cite{Tan}. The resulting embeddings are then trained on an MLP classifier. We were able to see a converging behavior of our model.

\begin{figure}[H]
    \hrule\vspace{0.1in}\centering
    \includegraphics[scale=0.45]{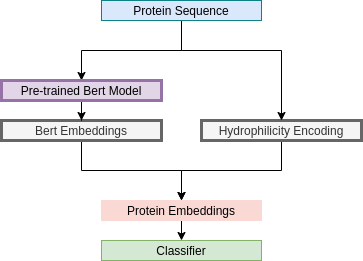}
    \vspace{0.1in}
    \hrule
    \caption{Model Architecture for Specification C}
    \label{fig:ArchitectureBertEmbedding}
\end{figure}

\begin{verbatim}
===========================================================================
Layer                Kernel Shape  Output Shape  Param #       Mult-Adds
===========================================================================
Sequential: 1-1                 -  [32,    9]            -                -
Linear: 2-1          [2048, 1024]  [32, 1024]    2,098,176       67,108,864
ReLU: 2-2                       -  [32, 1024]            -                -
Linear: 2-3           [1024, 512]  [32,  512]      524,800       16,777,216
ReLU: 2-4                       -  [32,  512]            -                -
Linear: 2-5            [512, 256]  [32,  256]      131,328        4,194,304
ReLU: 2-6                       -  [32,  256]            -                -
Linear: 2-7              [256, 9]  [32,    9]        2,313           73,728
===========================================================================
Total params:          2.756617M            Non-trainable params        0.0
Trainable params:      2.756617M            Total mult-adds (M):     881.54
===========================================================================
\end{verbatim}
\centerline{ Table 3: Model Metadata for Specification C}

\begin{figure}[H]
    \hrule\vspace{0.1in}
    \centering
    \includegraphics[scale=0.45]{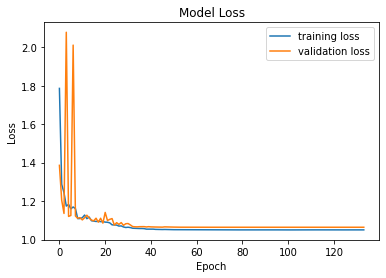}
    \includegraphics[scale=0.45]{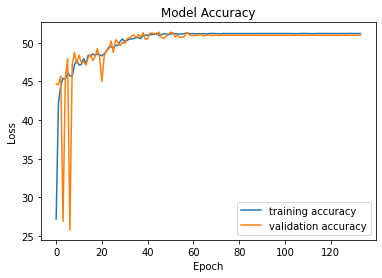}
    \vspace{0.1in}\hrule
    \caption{Accuracy and loss for Bert Embedding + Hydrophilicity Encoding}
    \label{fig:AccuracyLossBertEmbedding}
\end{figure}
\subsubsection{Hyperparameter Search for Specification C}
\begin{center}
    \begin{tabular}{||c c c c||} 
    \hline
    Initial LR & Weight Decay(Adam) & Training Acc & Validation Acc \\ [0.5ex] 
    \hline\hline
    1e-3 & 5e-3 & 51.5\% & 50.5\% \\ 
    \hline
    2e-3 & 5e-6 & 54.5\% & 53.5\% \\
    \hline
    3e-3 & 5e-7 & 54.7\% & 53.7\% \\
    \hline
    2e-3 & 1e-6 & 54.2\% & 52.3\% \\
    \hline
    3e-3 & 1e-3 & 55.2\% & 54.9\% \\
    \hline
    \end{tabular}
\end{center}

\subsection{Model Ablation Analysis}
In addition to the models presented above, we performed various experiments on the baseline model to determine whether each component in our network was necessary. Our results are summarized in the following table, and more detailed descriptions and results are available in Appendix II. In summary, while adding a heuristical embedding layer seemed to improve our baseline performance, this improvement was reversed by the other changes we tried such as removing dropout, removing convolutional layers, and removing LSTM layers:
\begin{center}
    \begin{tabular}{||l c c||} 
    \hline
    Architecture & Training Acc & Validation Acc \\ [0.5ex]
    \hline\hline
    Specification A: Baseline Model & 25.4\% & 24.9\% \\
    Specification D: Heuristical Embedding & 35.9\% & 36.2\% \\
    Specification E: Dropout Removed & 25.4\% & 25.3\% \\
    Specification F: CNN and Dropout Removed & 25.4\% & 25.6\% \\
    Specification G: Single LSTM Layer & 25.5\% & 25.7\% \\
    \hline
    \end{tabular}
\end{center}

\section{Understanding Model Behavior}
To verify how our model was learning, and debug any potential issues in our underlying implementation, we analyzed the gradient flow and outputted embeddings of our network.
\subsection{Gradient Flow Analysis}
To quantify whether our model was learning, we performed gradient flow analysis on our models by plotting the average and maximum gradient for each set of parameters in the network. The gradient flows are averaged over the first five epochs of training on the balanced dataset. We could observe that the input layers in the baseline LSTM are not being updated, while the input layers in the PLSTMs are having small updates each batch.

\begin{figure}[H]
    \hrule\vspace{0.1in}
    \centering
    \includegraphics[scale=0.45,valign=t]{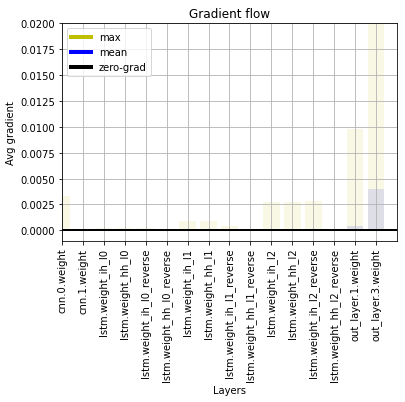}
    \includegraphics[scale=0.45,valign=t]{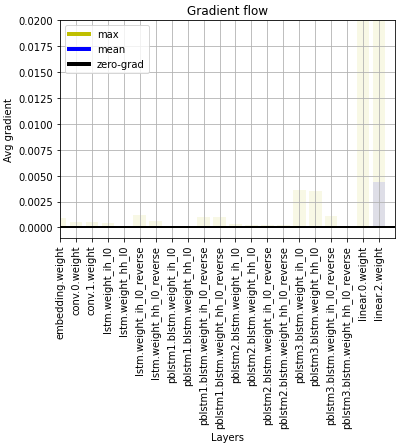}
    \includegraphics[scale=0.35]{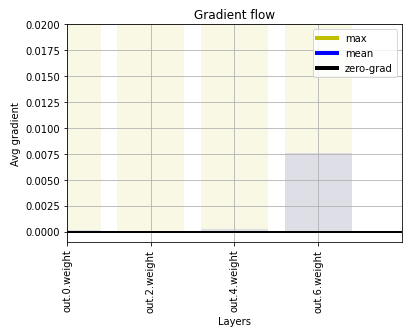}
    \vspace{0.1in}\hrule
    \caption{Gradient Flow for Baseline (top left), PLSTM (top right), and Bert Embedding + Hydrophilicity Encoding + MLP (bottom)}
    \label{fig:GradientFlow}
\end{figure}

The MLP model also suffers from vanishing gradients to some extent, since the average update on the weights in the first few linear layers is on the order of $10^{-4}$. This indicates that we may see a performance increase by adding more regularization techniques to the MLP model.

\subsection{Embedding Visualization}

To visualize whether our raw numerical sequence data and Bert embedding data are distinguishable from each other, we used T-SNE\footnote{\url{https://scikit-learn.org/stable/modules/generated/sklearn.manifold.TSNE.html}} from the scikit-learn package to plot the embeddings. More specifically, 10\% of the data is sub-sampled from the balanced raw and embedding training datasets, with an equal amount of sequence per output clade label. Raw sequences are padded to the same length before performing T-SNE. The results are shown below:

\begin{figure}[H]
    \hrule\vspace{0.1in}
    \centering
    \includegraphics[scale=0.4]{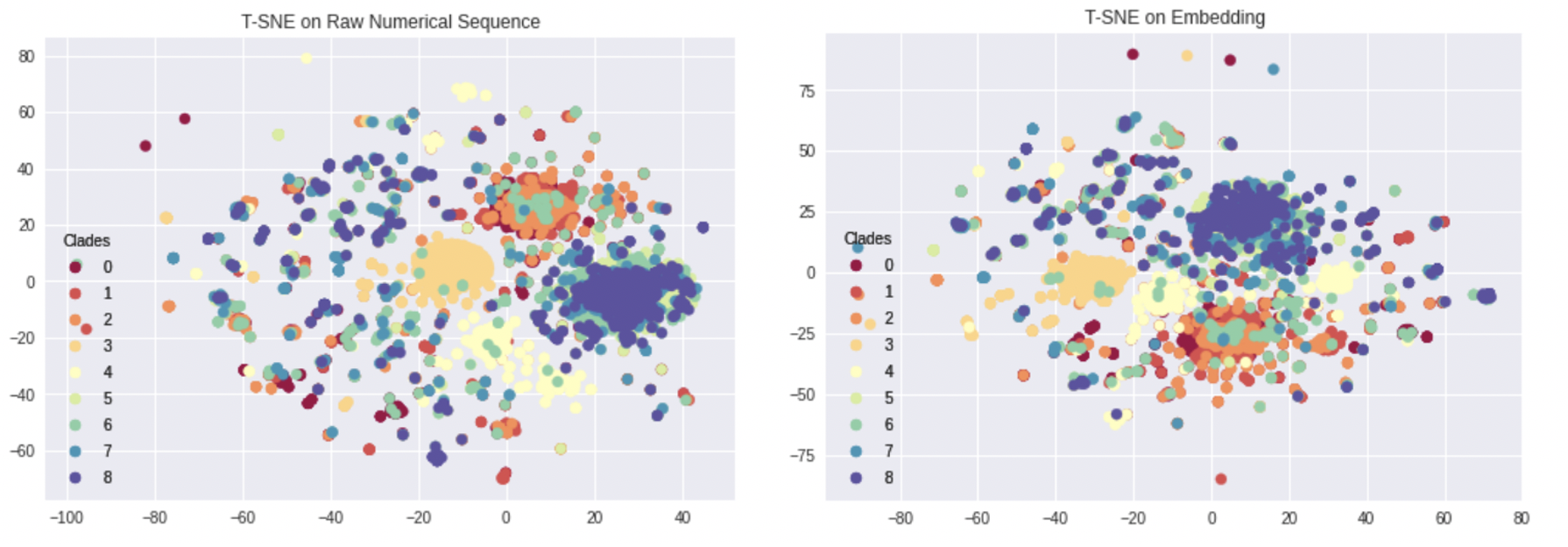}
    \vspace{0.1in}\hrule
    \caption{T-SNE Embedding Visualization}
    \label{fig:EmbeddingVisualization}
\end{figure}

While there are clades that form distinct clusters such as clades 3, 4 and 8 for both the raw and embedding data, most of the other clades' data points are rather spread out such as clades 0 and 5. The clades that do form distinct clusters are also the labels that our neural network model described above prefer to output. This visualization also explains why many of our models had difficulty differentiating among clades 0-2 and and clades 6-8 as these clusters overlap on the T-SNE plots.
Therefore, some of the clades are more similar to each other than others, and our naive numerical representation of amino acid sequence data might not be a good fit for this task. Using Bert embedding with hydrophillicity encoding does help to group data of the same clade together, but between group distances are reduced at the same time. 

\section{Future Works and Discussion}

\subsection{Major Challenges}
Because we were not able to find many existing papers that model our problem of classifying clades from protein sequencing data using neural networks, we lacked reference in how we should pre-process our data to account for all the nuances in biological data, as well as how to structure our networks. More specifically, our dataset is different from the frame-level phoneme classification task we were accustomed to because of the unique nature of protein folding - amino acids that are far downstream could have significant impacts on amino acids elsewhere (e.g. form disulfide bonds and other secondary or tertiary structures). Therefore, our LSTM model should not only learn to place an emphasis on nearby sequences, but also on amino acids much farther away which may have crucial impact on if a structural change in one site would lead to a clade change.

\subsection{Project Constraints}
Because we had limited computational resources and time for this project, we were not able to refine our model with much more advanced LSTM models or dedicate time finding better encoding methods since we spent much time refining and expanding our dataset in the early phase and had to re-run all our models each time we updated our dataset. For example, when we originally tried to run our baseline BiLSTM model on the full, unbalanced dataset containing over 700k sequences, it took us over two hours per epoch using GPU provided by Google Colab Pro. We also experimented with adding convolutional and fully connected layers to our baseline model. The maximum hidden size without triggering an OOM error is 256 for a second convolutional layer and 512 for a fully connected layers, which did not show a significant improvement over our previous result. Therefore, monetary and time cost limited us from experimenting more model structures. 

\subsection{Areas Needing Improvement}

\subsubsection{Data}
One limitation is that we are using spike protein sequences only due to limited computational power, while divergence labels are produced from the entire SARS-CoV-2 virus. We noticed that many of our sequences are the same despite having different divergence labels, and many other sequences differ at just one or two locations. Therefore, using the entire protein sequence could help resolve this problem and help the model to learn differences between sequences in relation with label.

Furthermore, when balancing our data, we randomly downsampled due to limited computational power, which could have lost some information contained in sequences belonging to groups that were downsampled. Oversampling and weighted penalty would be better approaches if we had better computational resources.

\subsubsection{Model}
Our pyramidal LSTM structure has shown improvements in helping the model to better condense and learn the input sequence; however, we are still yet to find embedding and encoding schemes that are based on biological heuristics and compatible with the model. More experiments should be done on how to combine the Bert embedding with a pyramidal LSTM architecture. Our model would also likely benefit from an attention mechanism to focus on key mutation sites, especially since the majority of amino acid sites in SARS-CoV-2 sequences are highly similar to each other.


\bibliographystyle{unsrt}  
\bibliography{ref}

\newpage
\appendix
\begin{center}
\Large{\bf{Appendix for Using Deep Learning Sequence Models to Identify SARS-CoV-2 Divergence}}
\end{center}

\section{Dataset Description}
In this section, we provide more details on our dataset, including example entries and histograms of the frequency of different clades and lineages.

Below is an example FASTA entry encoding a spike protein sequence:
\begin{verbatim}
{
    "Strain": "Wuhan/WIV04/2019",
    "Submission Date": "2019-12-30",
    "EPI_ISL": "EPI_ISL_402124",
    "Division of Exposure": "hCoV-19^^Hubei",
    "Originating Lab": "Wuhan Jinyintan Hospital",
    "Submitting Lab": "Wuhan Institute of Virology",
    "Author": "Shi",
    "Country of Exposure": "China",
    "Sequence": "MFVFLVLLPLVSSQCVNLTTRTQLPPAYTNSFTRGVYYPDKVFRSSVLHSTQDLFLPFF
        SNVTWFHAIHVSGTNGTKRFDNPVLPFNDGVYFASTEKSNIIRGWIFGTTLDSKTQSLLIVNNATNVV
        IKVCEFQFCNDPFLGVYYHKNNKSWMESEFRVYSSANNCTFEYVSQPFLMDLEGKQGNFKNLREFVFK
        NIDGYFKIYSKHTPINLVRDLPQGFSALEPLVDLPIGINITRFQTLLALHRSYLTPGDSSSGWTAGAA
        AYYVGYLQPRTFLLKYNENGTITDAVDCALDPLSETKCTLKSFTVEKGIYQTSNFRVQPTESIVRFPN
        ITNLCPFGEVFNATRFASVYAWNRKRISNCVADYSVLYNSASFSTFKCYGVSPTKLNDLCFTNVYADS
        FVIRGDEVRQIAPGQTGKIADYNYKLPDDFTGCVIAWNSNNLDSKVGGNYNYLYRLFRKSNLKPFERD
        ISTEIYQAGSTPCNGVEGFNCYFPLQSYGFQPTNGVGYQPYRVVVLSFELLHAPATVCGPKKSTNLVK
        NKCVNFNFNGLTGTGVLTESNKKFLPFQQFGRDIADTTDAVRDPQTLEILDITPCSFGGVSVITPGTN
        TSNQVAVLYQDVNCTEVPVAIHADQLTPTWRVYSTGSNVFQTRAGCLIGAEHVNNSYECDIPIGAGIC
        ASYQTQTNSPRRARSVASQSIIAYTMSLGAENSVAYSNNSIAIPTNFTISVTTEILPVSMTKTSVDCT
        MYICGDSTECSNLLLQYGSFCTQLNRALTGIAVEQDKNTQEVFAQVKQIYKTPPIKDFGGFNFSQILP
        DPSKPSKRSFIEDLLFNKVTLADAGFIKQYGDCLGDIAARDLICAQKFNGLTVLPPLLTDEMIAQYTS
        ALLAGTITSGWTFGAGAALQIPFAMQMAYRFNGIGVTQNVLYENQKLIANQFNSAIGKIQDSLSSTAS
        ALGKLQDVVNQNAQALNTLVKQLSSNFGAISSVLNDILSRLDKVEAEVQIDRLITGRLQSLQTYVTQQ
        LIRAAEIRASANLAATKMSECVLGQSKRVDFCGKGYHLMSFPQSAPHGVVFLHVTYVPAQEKNFTTAP
        AICHDGKAHFPREGVFVSNGTHWFVTQRNFYEPQIITTDNTFVSGNCDVVIGIVNNTVYDPLQPELDS
        FKEELDKYFKNHTSPDVDLGDISGINASVVNIQKEIDRLNEVAKNLNESLIDLQELGKYEQYIKWPWY
        IWLGFIAGLIAIVMVTIMLCCMTSCCSCLKGCCSCGSCCKFDEDDSEPVLKGVKLHYT*"
}
\end{verbatim}

Below is an example metadata entry encoding the mutations, clade, and pango lineage of a viral sequence:
\begin{verbatim}
{
    "Virus name": "Australia/NT12/2020",
    "Type": "betacoronavirus",
    "Accession ID": "EPI_ISL_426900",
    "Collection date": "2020",
    "Location": "Oceania / Australia / Northern territory",
    "Additional location information": "",
    "Sequence length": "29862",
    "Host": "Human",
    "Patient age": "unknown",
    "Gender": "unknown",
    "Clade": "G",
    "Pango lineage": "B.1",
    "Pangolin version": "2021-04-21",
    "Variant": "",
    "AA Substitutions": "(NSP15_A283V, NSP12_P323L, Spike_D614G)",
    "Submission date": "2020-04-17",
    "Is reference?": "",
    "Is complete?": "True",
    "Is high coverage?": "True",
    "Is low coverage?": "",
    "N-Content": "0.00691236470311",
    "GC-Content": "0.379674275888"
}
\end{verbatim}

Below is an example matched entry after correspondences are computed between protein sequences and metadata. The sequence is identical to the example FASTA entry shown above:
\begin{verbatim}
{
    "strain": "Australia/NT12/2020",
    "clade": "G",
    "pango_lineage": "B.1",
    "sequence": "MFVFLVLLPLVSSQCVNLTT...KFDEDDSEPVLKGVKLHYT*"
}
\end{verbatim}

Our amino acid mapping is as follows:
\begin{verbatim}
    {"A": 0, "B": 1, "C": 2, "D": 3, "E": 4, "F": 5, "G": 6, "H": 7, "I": 8,
     "K": 9, "L": 10, "M": 11, "N": 12, "P": 13, "Q": 14, "R": 15, "S": 16,
     "T": 17, "U": 18, "V": 19, "W": 20, "X": 21, "Y": 22, "Z": 23, "*": 24}
\end{verbatim}

Our clade mapping is as follows:
\begin{verbatim}
    {"G": 0, "GH": 1, "GR": 2, "GRY": 3, "GV": 4, "L": 5, "O": 6, "S": 7, "V": 8}
\end{verbatim}

We also produced a mapping from Pango lineage to integers. Because there are 1271 classes for Pango lineage labels, which is too many to show, we will just show the first and last 10 entries of this mapping:
\begin{verbatim}
    {"A": 0, "A.1": 1, "A.11": 2, "A.12": 3, "A.15": 4, "A.16": 5, "A.17": 6, "A.18": 7,
     "A.19": 8, "A.2": 9, ..., "U.2": 1261, "U.3": 1262, "V.1": 1263, "V.2": 1264,
     "W.1": 1265, "W.2": 1266, "W.3": 1267, "W.4": 1268, "Y.1": 1269, "Z.1": 1270}
\end{verbatim}

\clearpage

\section{Model Ablation Results}
In this section, we expand on the model ablation results summarized in Section 4.6. All these experiments are based on our baseline model architecture.

\subsection{Heuristical Embedding Layer Result: Specification D}

To provide a more meaningful input to the LSTM, we introduced a heuristical, learned embedding layer before the LSTM. Although these embeddings have no explicit biological meaning and are learned completely by our network, our hope is that they allow future layers of the LSTM to better represent the input amino acids. Our model description and torchsummary is as follows:

\begin{figure}[H]
    \hrule\vspace{0.1in}\centering
    \includegraphics[scale=0.45]{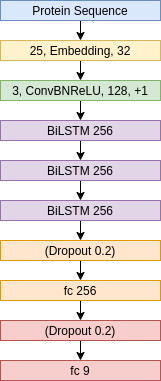}
    \vspace{0.1in}
    \hrule
    \caption{Model Architecture for Specification D}
    \label{fig:ArchitectureEmbedding}
\end{figure}

\begin{verbatim}
=============================================================================
                       Kernel Shape       Output Shape     Params   Mult-Adds
Layer                                                                      
0_embedding                [32, 25]  [1274,  32,   32]      800.0       800.0
1_conv.Conv1d_0        [32, 128, 3]  [  32, 128, 1274]    12.288k  15.654912M
2_conv.BatchNorm1d_1          [128]  [  32, 128, 1274]      256.0       128.0
3_conv.ReLU_2                     -  [  32, 128, 1274]          -           -
4_lstm                            -  [1274,  32,  512]  3.944448M    3.93216M
5_linear.Dropout_0                -  [1274,  32,  512]          -           -
6_linear.Linear_1        [512, 256]  [1274,  32,  256]   131.328k    131.072k
7_linear.Dropout_2                -  [1274,  32,  256]          -           -
8_linear.Linear_3          [256, 9]  [1274,  32,    9]     2.313k      2.304k
9_linear.LogSoftmax_4             -  [1274,  32,    9]          -           -
-----------------------------------------------------------------------------
Total params           4.091433M            Non-trainable params         0.0
Trainable params       4.091433M            Mult-Adds             19.721376M
=============================================================================
\end{verbatim}
\centerline{ Table 4: Model Metadata for Specification D}

Our results are as follows:

\begin{figure}[H]
    \hrule\vspace{0.1in}\centering
    \includegraphics[scale=0.45]{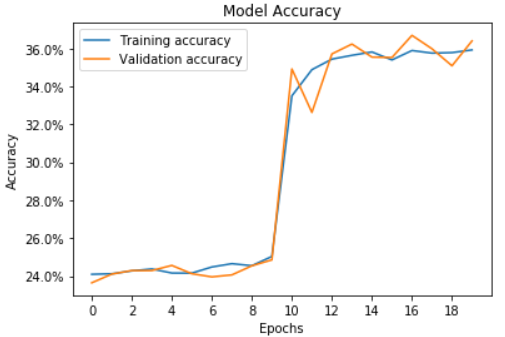}
    \includegraphics[scale=0.45]{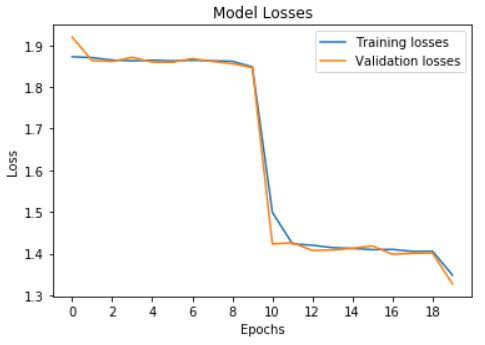}
    \vspace{0.1in}\hrule
    \caption{Accuracy and loss for LSTM model with embedding}
    \label{fig:AccuracyLossEmbedding}
\end{figure}

We also plotted a histogram of model output frequencies vs. label frequencies, as well as a confusion matrix describing the classification error of the model:
\begin{center}
    \hrule\vspace{0.1in}
    \includegraphics[scale=0.45]{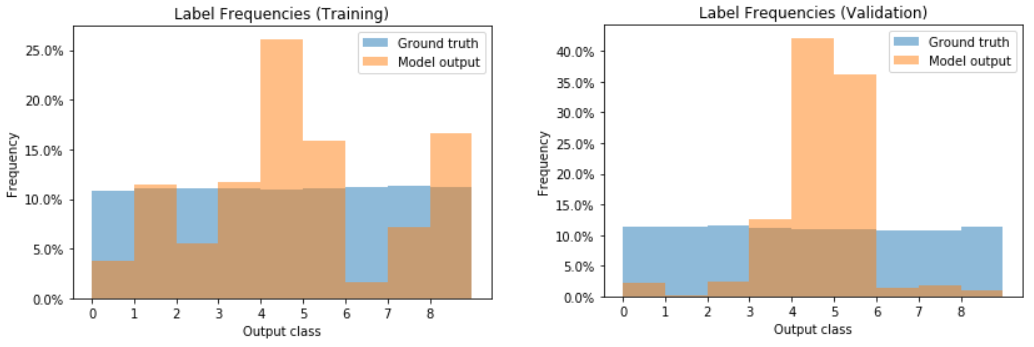}
\end{center}
\begin{figure}[H]
    \centering
    \includegraphics[scale=0.45]{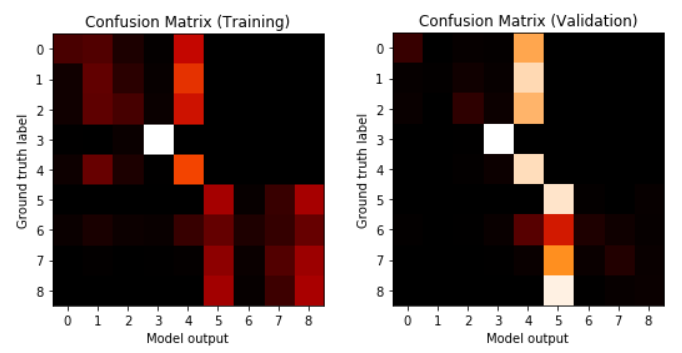}
    \vspace{0.1in}\hrule
    \caption{Output label frequencies and confusion matrix for LSTM model with embedding}
    \label{fig:FrequencyConfusionEmbedding}
\end{figure}

It seems that introducing a heuristical embedding layer slightly improved our model's performance, as our model is now able to correctly identify clades 3, 4, and 5 after training for some epochs. However, our model continues to struggle with identifying the remaining clades. Clades 0, 1, and 2 seem to get confidently identified as clade 4, and clades 6, 7, and 8 seem to get confidently identified as clade 5.

\clearpage

\subsection{Baseline with Dropout Removed: Specification E}

Since our model is not currently performing very well, we progressively reduced our model complexity to determine whether each component of our model was needed, and to verify what kind of results we could produce from the simplest of LSTM architectures.

We began by removing dropout on the final linear layers, because dropout is typically to perform model regularization but may make it more difficult for a model to learn. Our model description and torchsummary is as follows:

\begin{figure}[H]
    \hrule\vspace{0.1in}\centering
    \includegraphics[scale=0.45]{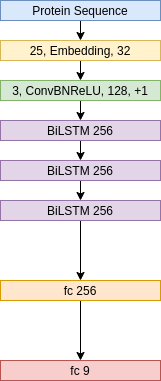}
    \vspace{0.1in}
    \hrule
    \caption{Model Architecture for Specification E}
    \label{fig:ArchitectureNoDropout}
\end{figure}

\begin{verbatim}
=============================================================================
                       Kernel Shape       Output Shape     Params   Mult-Adds
Layer                                                                      
0_embedding                [32, 25]  [1274,  32,   32]      800.0       800.0
1_conv.Conv1d_0        [32, 128, 3]  [  32, 128, 1274]    12.288k  15.654912M
2_conv.BatchNorm1d_1          [128]  [  32, 128, 1274]      256.0       128.0
3_conv.ReLU_2                     -  [  32, 128, 1274]          -           -
4_lstm                            -  [1274,  32,  512]  3.944448M    3.93216M
5_linear.Linear_0          [512, 9]  [1274,  32,    9]     4.617k      4.608k
6_linear.LogSoftmax_1             -  [1274,  32,    9]          -           -
-----------------------------------------------------------------------------
Total params           3.962409M            Non-trainable params         0.0
Trainable params       3.962409M            Mult-Adds             19.592608M
=============================================================================
\end{verbatim}

\centerline{ Table 5: Model Metadata for Specification E}

Our results are as follows:
\begin{figure}[H]
    \hrule\vspace{0.1in}\centering
    \includegraphics[scale=0.45]{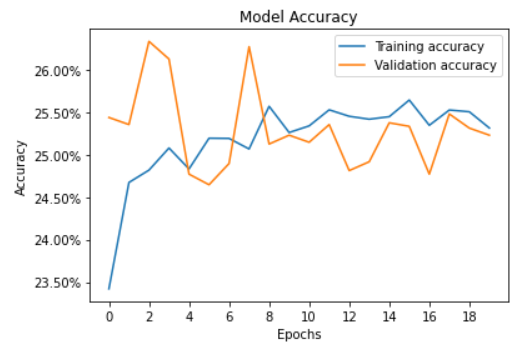}
    \includegraphics[scale=0.45]{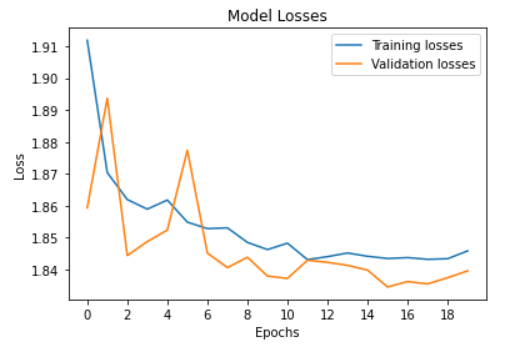}
    \vspace{0.1in}\hrule
    \caption{Accuracy and loss for LSTM model without dropout}
    \label{fig:AccuracyLossNoDropout}
\end{figure}

We also plotted a histogram of model output frequencies vs. label frequencies, as well as a confusion matrix describing the classification error of the model:
\begin{center}
    \hrule\vspace{0.1in}
    \includegraphics[scale=0.45]{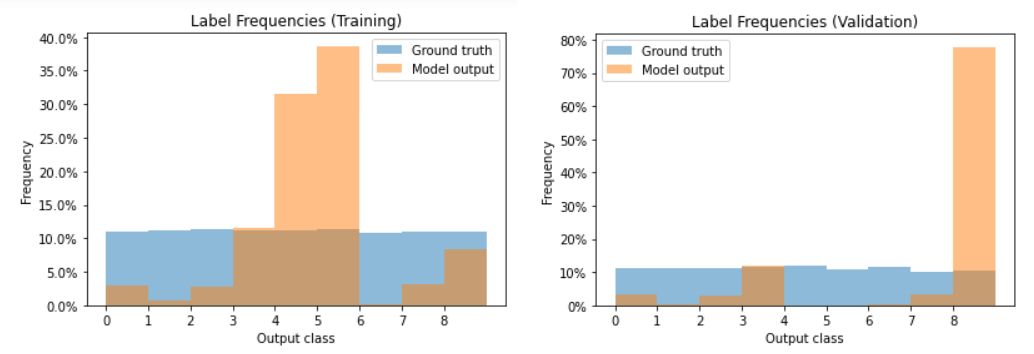}
\end{center}
\begin{figure}[H]
    \centering
    \includegraphics[scale=0.45]{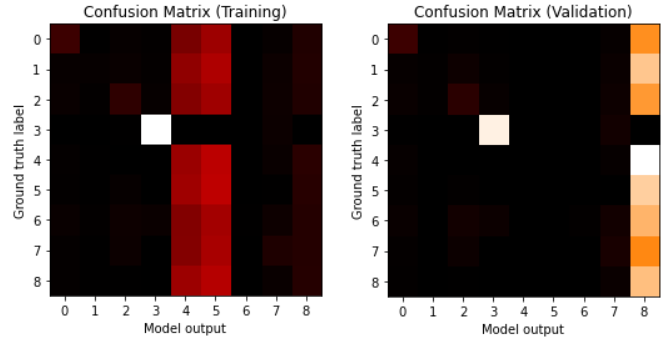}
    \vspace{0.1in}\hrule
    \caption{Output label frequencies and confusion matrix for LSTM model without dropout}
    \label{fig:FrequencyConfusionNoDropout}
\end{figure}

While removing dropout seems to very slightly improve our model performance at the start by less than 1\%, the model seems to lose its ability to identify clades 4 and 5, as that ability was gained at epoch 10 when dropout was included but was never gained when dropout was removed. Although this increase in performance is unlikely due to the regularization benefit that dropout provides, introducing dropout also has other effects such as encouraging hidden unit activations to become more sparse \cite{Srivastava}, which may influence the underlying representation and performance of our model.

\clearpage

\subsection{Baseline with CNN and Dropout Removed: Specification F}

As a next step, we removed the convolutional layers before the LSTM layers, since convolutional layers before LSTM are typically used to improve the quality of embeddings passed into the LSTM, but they add an extra degree of complexity and may not be strictly necessary for optimal performance. Our model description and torchsummary is as follows:

\begin{figure}[H]
    \hrule\vspace{0.1in}\centering
    \includegraphics[scale=0.45]{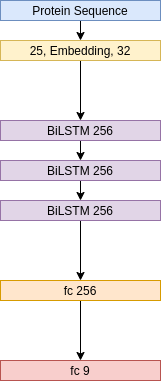}
    \vspace{0.1in}
    \hrule
    \caption{Model Architecture for Specification F}
    \label{fig:ArchitectureNoConv}
\end{figure}

\begin{verbatim}
==========================================================================
                      Kernel Shape       Output Shape    Params  Mult-Adds
Layer                                                                   
0_embedding               [32, 25]  [1274,  32,   32]     800.0      800.0
1_lstm                           -  [1274,  32,  512]  3.74784M  3.735552M
2_linear.Dropout_0               -  [1274,  32,  512]         -          -
3_linear.Linear_1       [512, 256]  [1274,  32,  256]  131.328k   131.072k
4_linear.Dropout_2               -  [1274,  32,  256]         -          -
5_linear.Linear_3         [256, 9]  [1274,  32,    9]    2.313k     2.304k
6_linear.LogSoftmax_4            -  [1274,  32,    9]         -          -
--------------------------------------------------------------------------
Total params          3.882281M            Non-trainable params        0.0
Trainable params      3.882281M            Mult-Adds             3.869728M
==========================================================================
\end{verbatim}
\centerline{ Table 5: Model Metadata for Specification F}
Our results are as follows:
\begin{figure}[H]
    \hrule\vspace{0.1in}\centering
    \includegraphics[scale=0.45]{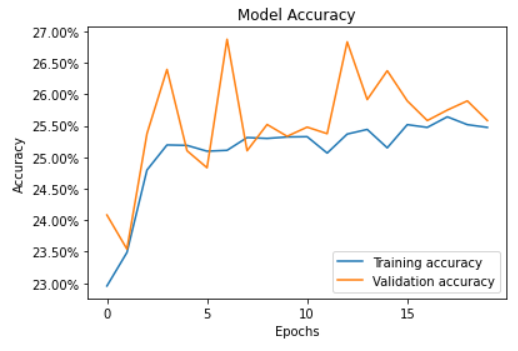}
    \includegraphics[scale=0.45]{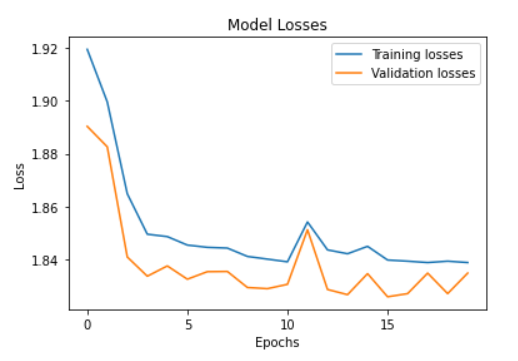}
    \vspace{0.1in}\hrule
    \caption{Accuracy and loss for LSTM model without convolutional layers}
    \label{fig:AccuracyLossNoConv}
\end{figure}

We also plotted a histogram of model output frequencies vs. label frequencies, as well as a confusion matrix describing the classification error of the model:
\begin{center}
    \hrule\vspace{0.1in}
    \includegraphics[scale=0.45]{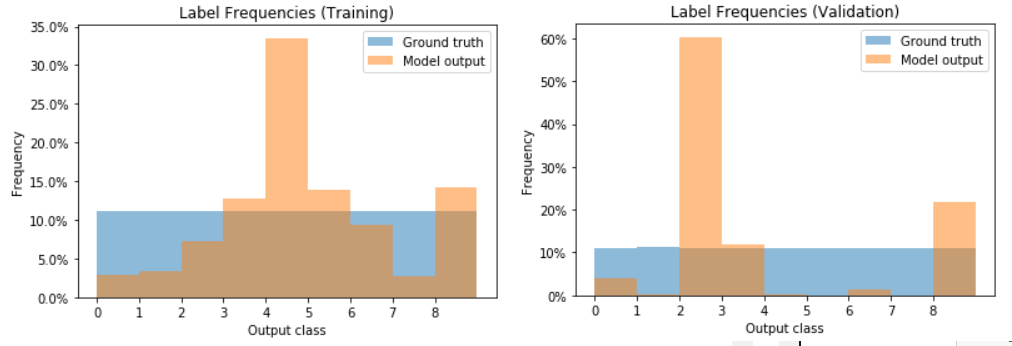}
\end{center}
\begin{figure}[H]
    \centering
    \includegraphics[scale=0.45]{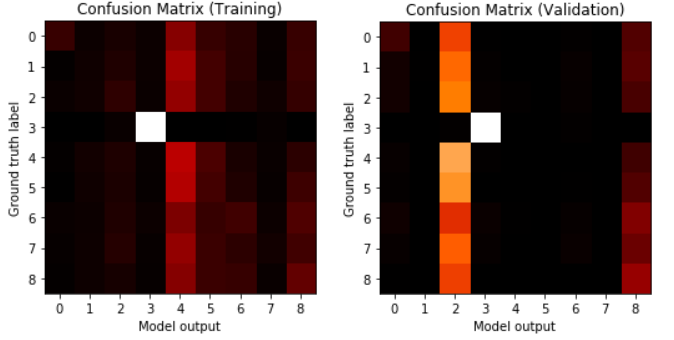}
    \vspace{0.1in}\hrule
    \caption{Output label frequencies and confusion matrix for LSTM model without convolution}
    \label{fig:FrequencyConfusionNoConv}
\end{figure}

Removing the initial convolutional layers seems to limit our performance as well, although less so than removing dropout. Our model can reliably identify clade 3, which all our other models are able to do, but the model can no longer confidently identify any of the other clades. However, our model does learn to have some understanding of each clade, as evidenced by the slight diagonal in the training confusion matrix. 

\clearpage

\subsection{Single LSTM Layer: Specification G}

Finally, we stripped the model down to the absolute minimum of a single LSTM layer followed by a fully connected layer, to determine how the number of layers impacts model performance. Our model description and torchsummary is as follows:

\begin{figure}[H]
    \hrule\vspace{0.1in}\centering
    \includegraphics[scale=0.45]{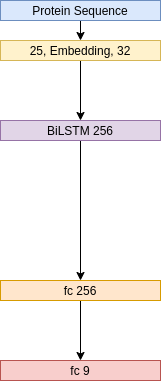}
    \vspace{0.1in}
    \hrule
    \caption{Model Architecture for Specification G}
    \label{fig:ArchitectureOneLayer}
\end{figure}

\begin{verbatim}
========================================================================
                      Kernel Shape     Output Shape    Params Mult-Adds
Layer                                                                  
0_embedding               [32, 25]   [1274, 32, 32]     800.0     800.0
1_lstm                           -  [1274, 32, 512]   593.92k  589.824k
2_linear.Dropout_0               -  [1274, 32, 512]         -         -
3_linear.Linear_1       [512, 256]  [1274, 32, 256]  131.328k  131.072k
4_linear.Dropout_2               -  [1274, 32, 256]         -         -
5_linear.Linear_3         [256, 9]    [1274, 32, 9]    2.313k    2.304k
6_linear.LogSoftmax_4            -    [1274, 32, 9]         -         -
------------------------------------------------------------------------
                        Totals
Total params          728.361k            Non-trainable params       0.0
Trainable params      728.361k            Mult-Adds               724.0k
========================================================================
\end{verbatim}

\centerline{ Table 6: Model Metadata for Specification G}
Our results are as follows:
\begin{figure}[H]
    \hrule\vspace{0.1in}\centering
    \includegraphics[scale=0.45]{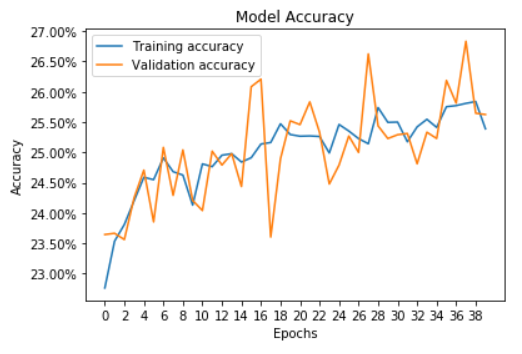}
    \includegraphics[scale=0.45]{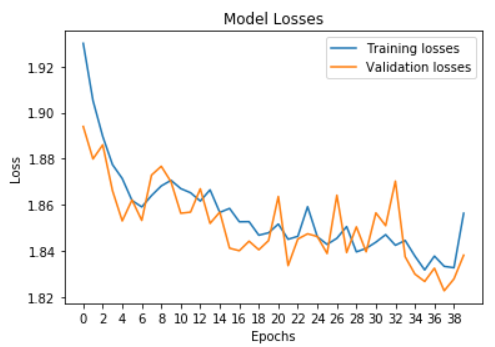}
    \vspace{0.1in}\hrule
    \caption{Accuracy and loss for LSTM model with one layer}
    \label{fig:AccuracyLossOneLayer}
\end{figure}

We also plotted a histogram of model output frequencies vs. label frequencies, as well as a confusion matrix describing the classification error of the model:
\begin{center}
    \hrule\vspace{0.1in}
    \includegraphics[scale=0.45]{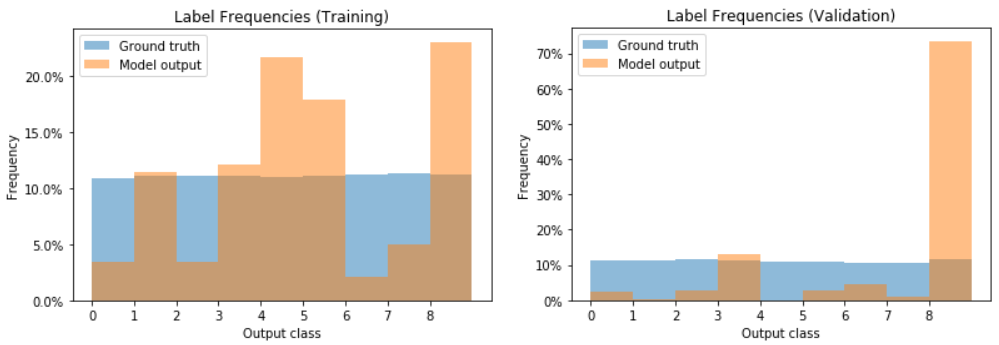}
\end{center}
\begin{figure}[H]
    \centering
    \includegraphics[scale=0.45]{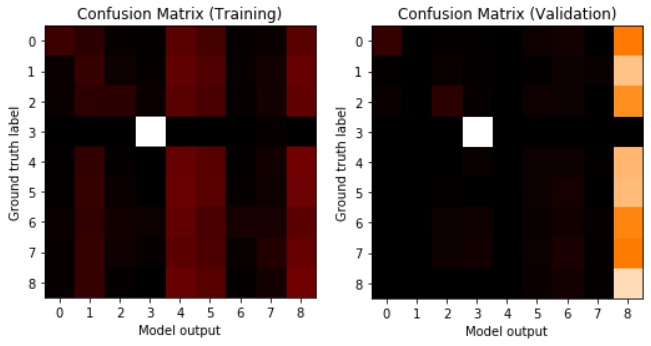}
    \vspace{0.1in}\hrule
    \caption{Output label frequencies and confusion matrix for LSTM model with one layer}
    \label{fig:FrequencyConfusionOneLayer}
\end{figure}

It seems our model's results are more unstable with one layer, and while a minimal LSTM model retains the ability to identify clade 3, it continues to do poorly at identifying the other clades. Our model also seems to have no understanding of clades 4-8, as those clades are depicted as straight vertical lines on the confusion matrix.

\end{document}